\documentclass[prl,showpacs,twocolumn,floatfix,unsortedaddress,superscriptaddress]{revtex4}
\usepackage{graphicx}

\def\bea{\begin{eqnarray}}
\def\eea{\end{eqnarray}}
\def\ben{\begin{equation}}
\def\een{\end{equation}}
\def\benu{\begin{enumerate}}
\def\enu{\end{enumerate}}

\def\n{n}

\def\sss{\scriptscriptstyle\rm}

\def\1var{(\bx_1...\bx\N)}

\def\br{{\bf r}}

\def\bx{{x}}

\def\bq{{\bf q}}

\def\x{_{\sss X}}
\def\c{_{\sss C}}
\def\s{_{\sss S}}
\def\xc{_{\sss XC}}

\def\N{_{\sss N}}

\def\GGA{^{\rm GGA}}

\def\unif{^{\rm unif}}
\def\surf{^{\rm surf}}
\def\excsurf{e\xc\surf}
\def\exsurf{e\x\surf}

\def\sph_int{ {\int d^3 r}}

\input rotate

\begin{document}
\def\z{_\zeta}
\def\GE{_{GE}}
%\large
%\sf
\def\x{_x}
\def\xc{_{xc}}
\def\c{_c}
\def\X{$x~$}
\def\XC{$xc~$}
\def\C{$c~$}

\title{Restoring the 
density-gradient expansion for exchange in solids and surfaces}
\author{John P. Perdew}
\author{Adrienn Ruzsinszky}
\affiliation{Department of Physics and Quantum Theory Group, Tulane University, New Orleans, LA 70118}
\author{G\'abor I. Csonka}
\affiliation{Department of Chemistry, Budapest U. of Technology and Economics, H-1521                         
Budapest, Hungary}
\author{Oleg A. Vydrov}
\author{Gustavo E. Scuseria}
\affiliation{Department of Chemistry, Rice University, Houston, TX 77005}
\author{Lucian A. Constantin}
\affiliation{Donostia International Physics Center, E-20018, Donostia, Basque Country}
\author{Xiaolan Zhou}
\affiliation{Department of Physics and Quantum Theory Group, Tulane University, New Orleans, LA 70118}
\author{Kieron Burke}
\affiliation{Departments of Chemistry and of Physics, UC Irvine, CA 92697}
\date{\today}
\begin{abstract} 
Successful modern generalized gradient approximations (GGA's) are
biased toward atomic energies. Restoration of the first-principles gradient
expansion for exchange over a wide range of density gradients
eliminates this bias.
We introduce PBEsol, a revised 
Perdew-Burke-Ernzerhof GGA that improves equilibrium properties of 
densely-packed solids and their surfaces.
\end{abstract}


\maketitle
%\Large\sf

%%%%%%%%%%%%%%%%%%%%%%%%%%%%%%%%%%%%%%
%%%%%%%%%%% INTRODUCTION %%%%%%%%%%%%%
%%%%%%%%%%%%%%%%%%%%%%%%%%%%%%%%%%%%%%

Ground-state Kohn-Sham density functional theory (DFT) 
\cite{KS65}
has been hugely successful
for electronic structure calculations of solids
and molecules.  Its accuracy relies upon good
approximations to the unknown exchange-correlation
(\XC) energy as a functional of the electronic spin densities\cite{FNM03}.
Over the past four decades, a variety of increasingly
sophisticated approximations has been developed\cite{TPSS03}.
The most commonly used in solid-state calculations
today is the PBE version\cite{PBE96} of the generalized gradient
approximation (GGA), employing both the density and its
gradient at each point in space.
Popular GGAs represent a well-tempered balance
between computational efficiency, numerical accuracy,
and reliability, but PBE also juggles the demands of
quantum chemistry and solid-state physics\cite{KPB99}.

While PBE represented a high-point of non-empirical functional
development 11 years ago, 
much has since been learned about its limitations.
PBE reduces the chronic overbinding of 
the local
spin density approximation (LSDA)\cite{KS65} but,
while LSDA
often  slightly underestimates equilibrium
lattice constants by about 1\%, PBE usually overestimates them
by about the same amount.  Other equilibrium properties, such
as bulk moduli, phonon frequencies, magnetism, and ferro-electricity,
are 
sensitive to the lattice constant, and so are also overcorrected
by PBE\cite{WC06}.   Surface energies are
too low in LSDA, but are made lower still by PBE\cite{CPT06}.

However, attempts to construct a better GGA face a Procrustean
dilemma\cite{PBE98}:
Those with an enhanced gradient dependence\cite{ZY98,HHN99} improve
atomization and total energies, but worsen bond lengths, while
more recent
suggestions 
of a GGA for solids and surfaces\cite{WC06,VJKS00,AM05,TLBS07,VRRK07,CVSR07}
have a reduced gradient dependence and typically do improve lattice
parameters and/or surface energies, but have been criticized for worsening
energetics\cite{TLBS07}.
More advanced functionals have been constructed.
For example, meta-GGA's, using also the orbital kinetic-energy 
densities, provide greater accuracy over a wider range of systems
and properties\cite{TPSS03}.
But current meta-GGA's do not improve lattice constants as
dramatically as surface energies, and
meta-GGA's are not yet available in all solid-state codes.

In the present work, we explain the origin of this dilemma and show
that, in a sense, {\em no} GGA can do both:  Accurate atomic
exchange energies require violating the gradient expansion for slowly-varying
densities, which is valid for solids and their surfaces.  At the GGA level,
one must choose.  
A pragmatic approach to lattice properties is therefore to
use a modified functional especially for solids which, unlike
previous suggestions, recovers the
gradient expansion for exchange.  Such a functional becomes exact
for solids under intense
compression.
Analogous reasoning for correlation suggests fitting the jellium \XC surface
energy, as was first done by Armiento and Mattsson\cite{AM05}.
Our variation on PBE satisfies these
new conditions, by simply
altering two parameters in the original formula.   
We find our new functional
significantly improves most lattice constants and surface
energies while worsening
atomization energies
(as does AM05
of Ref. \cite{AM05}).
By restoring the gradient expansion, PBEsol also yields 
excellent jellium surface exchange
energies, and provides an improved
starting point for more advanced functional construction.

The GGA form for the exchange energy is simply
\ben
E\x\GGA[\n] = \int d^3r\, e\x\unif(\n(\br))\, F\x(s(\br))
\een
where $\n(\br)$ is the electronic density, $e\x\unif(n)$ is 
the exchange energy density of a uniform electron gas (proportional to
$n^{4/3}$), $s=|\nabla \n|/(2 k_F \n)$ (with $k_F=
(3\pi^2\n)^{1/3}$) is the dimensionless density gradient, and
$F\x(s)$ is the so-called enhancement factor for the given GGA\cite{PBE96}.
Eq. (1) is the spin-unpolarized form, from which 
the spin dependence can be deduced\cite{FNM03}.
Any GGA that recovers the uniform gas limit has
\ben
F\x(s) = 1 + \mu\, s^2 + \ldots.~~~~~(s\to 0)
\een
The gradient expansion\cite{KS65}
that is accurate for slowly-varying electron gases has\cite{AK85}
$\mu\GE =10/81 \approx  .1234$.

To begin,
Ref \cite{PCSB06} showed that the
exchange energies of neutral atoms are very well approximated
by their asymptotic expansion for large $Z$, i.e., $E\x =
-.2208 Z^{5/3} -.196 Z +...$.   The first term arises from
LSDA, but the second arises in a GGA from the $s^2$
contribution to Eq. (2) and requires $\mu\approx 2 \mu\GE$.
Thus any GGA that is accurate for the exchange energies of neutral
atoms must have $\mu\approx 2\mu\GE$.
PBE does, although its value 
of $\mu=0.219$ was
found from a different non-empirical argument.  So does B88, as it
was fitted to the \X energies of noble gas atoms\cite{B88}.  Even
PW91 essentially does too\cite{PBW96}, as only
at irrelevantly small values of $s$ does it revert to $\mu\GE$.

%But $s \lesssim 1$ for valence electrons in densely-packed
%solids (or $s \lesssim 2$ in core-valence regions of
%alkali atoms), as is the reduced Laplacian,
%so the gradient expansion is more likely to be relevant here.
%If we remove the tyrannical yoke of reproducing
%cohesive energies in solids and atomization energies
%in molecules, we should restore $\mu=\mu\GE$.

%  If we    
%do not insist upon accurate cohesive energies for solids or atomization energies for molecules, we
%should restore $\mu=\mu_{GE}$.  

Thus, to attain accurate exchange energies
of atoms (vital to dissociation
energies in molecules and cohesive energies in solids),
any GGA must strongly violate the gradient expansion for
slowly-varying densities\cite{PCSB06}.
But most of thermochemistry occurs without free atoms, and is not much worse
in LSDA than in PBE (e.g., \cite{Bb06}). 
Moreover, for the evaluation of exchange, the
densities of real solids and their surfaces are often
almost slowly-varying over space.  Restoring the gradient
expansion should improve their description (but 
worsen atomization energies).  The GGA is a limited form, and {\em
cannot} satisfy both conditions.  
Eq. (2) suggests a necessary condition for convergence of the second-order          
gradient expansion for exchange: $s \lesssim 1$.  Since 
$s \lesssim 1$  for valence electrons in densely-packed 
solids (or $s \lesssim 2$ in core-valence regions of alkali atoms),
and since the reduced Laplacian of the
density is also $\lesssim 1$, the gradient expansion is
important for exchange in solids.
We choose $\mu\GE$ for PBEsol.

Now, for a GGA correlation functional that recovers the uniform gas limit,
the gradient expansion is 
\ben
E\c[\n] = \int d^3r\, \n(\br)\, \{ \epsilon\c\unif(\n(\br))+\beta\, t^2(\br)\, + \ldots\}
\een
where
$ \epsilon\c\unif(\n)$ is the correlation energy per particle of the
uniform gas, $\beta$ is a coefficient, and 
$t =|\nabla n|/\{2\, k_{TF}\, \n\}$ is the appropriate
reduced density gradient for correlation
(fixed by the Thomas-Fermi screening
wavevector $k_{TF}={\sqrt{4 k_F/\pi}}$,
not $k_F$.)
For slowly-varying high densities\cite{MB68},
$\beta\GE=0.066725$.
Unlike exchange, the second-order term in      
the gradient expansion for correlation cannot be small
 compared to the local term everywhere even for valence electrons in solids:
 $\beta t^2$ can be large compared to $|\epsilon\c\unif|$
(as $\beta\GE t^2 = 0.1s^2/r\s$).  The gradient expansion
can be relevant to real systems (especially solids) for
exchange, but much less so for correlation.

The gradient expansion for correlation is less
relevant to solids than is 
$f\xc(q)$ for the response of the uniform gas
to 
a weak potential
$\lambda\, \cos (\bq\cdot\br)$.  The exact $f\xc(q)$ is
almost independent of $q$, up to
$2k_F$ \cite{MCS95}.
Thus LSDA, which produces a constant (the value at $q=0$),
yields an accurate approximation 
for $q \lesssim 2 k_F$.  But any GGA with a
non-zero \XC contribution to second order in $\nabla n$
produces a term quadratic in $q$.
Since we are interested in weakly-varying valence
electron densities in densely-packed solids, we wish to retain this
excellent feature of LSDA.  If
\ben
\mu = \pi^2 \beta/3,
\label{mubet}
\een
there is complete cancellation between beyond-LSDA \X and \C 
contributions, restoring LSDA response.

In PBE, the gradient expansion for correlation is respected, i.e.,
$\beta=\beta\GE$, and $\mu \approx 2\mu\GE$ satisfies Eq. (\ref{mubet}).
This choice agrees well
with
the PW91 exchange functional, and the numerical
real-space cutoff construction\cite{PBW96} of the GGA
exchange
energy,
and yields
highly accurate exchange energies of
atoms.
But we have already argued that $\mu \approx 2\mu\GE$ is harmful for
many condensed matter applications.
Once we choose $\mu\GE$ for exchange, we cannot 
recover simultaneously 
the GEA for correlation and the linear response of LSDA for a 
uniform density.  Exact satisfaction of Eq. (\ref{mubet}) would
yield $\beta=0.0375$, but an increased
value will satisfy another, more relevant constraint for solid-state
applications.

%Thus the set of
%solids to which out new GGA is applicable should include at least those with
%moderately-varying valence electron densities.
%%
%% Armiento and Mattsson\cite{AM05}  constructed a GGA using
%surface energies, and found improved lattice parameters, but worse
%atomic energies.  Wu and Cohen\cite{WC06} tried to fit a GGA form to the more 
%sophisticated TPSS meta-GGA\cite{TPSS03}, and found improved jellium surface
%energies and lattice constants, but a test of their functional by
%Tran et al\cite{TLBS07} showed that PBE remains superior for the energetics
%of covalent and non-covalent bonds.  Finally several versions of
%PBE with reduced gradient dependence have been proposed and
%found to improve properties of solids\cite{VRRK07,CVSR07}.

For correlation, large neutral jellium clusters are our paradigm,
for which
$E\xc \to e\xc\unif V + \excsurf A + ..$ as the radius grows,
where $\excsurf$ is the jellium surface \XC energy, $V$ the
volume of the cluster and $A$ its area.
A GGA that recovers $\excsurf$ will be 
correct in leading- and next-order for neutral jellium
clusters as $N\to\infty$, in a similar way to modern
exchange GGA's for neutral atoms.
Moreover, the surface energy is
dominated by \XC contributions and $\excsurf$ is
really a bulk-like property, arising mainly (103\% at $r\s=2$) from
a moderately-varying-density region (with $s \lesssim 1$) inside
the classical turning plane. 

We check that this condition is compatible with the restoration
of the gradient expansion for exchange.  Because jellium clusters have
a uniform bulk density and because
most of the surface energy comes from within, the
gradient expansion should be accurate.
We find, at bulk density $r\s=3$, the 
errors of the surface exchange energy are: LSDA 27\%, PBE -11\% and PBEsol 2.7\%.

\begin{figure}[htb]
\unitlength1cm
\begin{picture}(12.5,5.5)
\put(-7,9){\makebox(12,6.5){
\includegraphics{rat.psp}
}}
\end{picture}
\caption{Ratio of calculated surface exchange-correlation
to that of LSDA as a function of $r\s$
for various approximations.}
\end{figure}

We next follow the lead of Ref. \cite{AM05}, and fit
$\excsurf$ to determine our correlation functional.
But the jellium $\excsurf$ is not known exactly.
Figure 1 shows surface energy enhancements relative
to LSDA.
The likely ``range of the possible''         
for $\excsurf$ extends from TPSS
meta-GGA
\cite{TPSS03,CPT06} 
or RPA+
 \cite{YPK00}
at the low end of what is possible (in agreement with the most
recent Quantum Monte Carlo calculations\cite{WHFG07})
to the RPA-like Pitarke-Perdew (PP)\cite{PP03}
value at the high end.
The TPSS value      
probably provides the best value that a GGA should try to 
achieve.  We choose $\beta=0.046$
and $\mu=\mu\GE$ (within the PBE form) for PBEsol, to best fit the TPSS results.
PBEsol should improve most
surface energies over LSDA, whereas PBE worsens them.

Thus we have violated Eq. (\ref{mubet}) in favor
of good surface energies.
But our value for $\beta$ is considerably closer to that
of the linear response requirement (0.037) than that demanded by
complete restoration
of the gradient expansion (0.066).  The linear response of
PBEsol is reasonably close to that of LSDA, and closer than 
that of the gradient expansion.

\begin{figure}[htb]
\unitlength1cm
\begin{picture}(12.5,11)
\put(-7,6){\makebox(12,11){
\includegraphics{Fxc.psp}
}}
\end{picture}
\caption{Enhancement factors of PBE and PBEsol, for
spin-unpolarized systems, as a function of reduced density
gradient, for various values of $r\s$.}
\end{figure}
PBEsol is exact for the uniform gas, and highly accurate for large
jellium clusters, both of moderate and high density.  But
PBEsol also becomes
exact for solids under intense compression, where
real solids and their surfaces become truly 
slowly-varying, and exchange dominates over correlation\cite{PCSB06}.
In Fig. 2, we plot the enhancement factors of PBE and PBEsol.
For a spin-unpolarized ($\zeta=0$) density $\n = 3/(4\pi r\s^3)$, we define
$F\xc(r\s,s)$
by
\ben
E\xc\GGA[\n] = \int d^3r\, e\x\unif(\n(\br))\, F\xc(r\s(\br),s(\br))
\een
The high-density ($r\s\to 0$) limit is $F\x(s)$ of Eq. (1).
The nonlocality or
$s$-dependence of GGA exchange is diminished from
PBE to PBEsol, making the latter
somewhat closer to LSDA.
Over the whole range $s\lesssim 1$, the PBEsol
$F\x$ is close to $1+\mu\GE s^2$.
  The range
$0 \lesssim s \lesssim 3$
is energetically important for
most properties of most real systems, while
$0\lesssim s\lesssim 1$
and $1< r\s <10$ are the ranges for
valence-electron regions in many densely-packed solids.

\begin{table}[htb]
\begin{tabular}{|c|cccc|}
\hline
class&LSDA&PBE&TPSS&PBEsol\\
\hline
\multicolumn{5}{|c|}{mean error}\\
4 simple metals &-9.0&2.9&5.3&-0.3\\
5 semiconductors &-1.1&7.9&6.2&3.0\\
5 ionic solids &-8.4&8.5&6.8&2.0\\
4 transition metals &-4.0&6.4&2.5&0.0\\
total &-5.5&6.6&5.4&1.3\\
\hline
\multicolumn{5}{|c|}{mean absolute error}\\
4 simple metals &9.0&3.4&5.3&2.3\\
5 semiconductors &1.3&7.9&6.2&3.0\\
5 ionic solids &8.4&8.5&6.8&2.7\\
4 transition metals &4.0&6.4&2.7&1.9\\
total &5.6&6.7&5.4&2.5\\
\hline
\end{tabular}
\label{t:G2}
\caption{Errors in equilibrium lattice constants (in \AA$\times 10^{-2}$)
on our data set of 18 solids, relative to experiment
with estimates of the zero-point anharmonic 
expansion removed\cite{SSTP04}.}
\end{table}
To test our functional, we employ a test set of 18 solids from
Ref \cite{SSTP04}.  
These come in four groups:  Simple metals (Li,Na,K,Al),
semiconductors (C,Si,SiC,Ge,GaAs), ionic solids (NaF,NaCl,LiCl,LiF,MgO),
and transition metals (Cu,Rh,Pd,Ag).  The set is not claimed to be 
representative, but was chosen for the availability of basis functions
and anharmonic corrections\cite{SSTP04}.
Our calculations use the 
Gaussian orbital periodic code of Ref \cite{SSTP04},
with basis sets of the same or higher 
quality.
In Table I, we list both the mean errors and the mean absolute errors 
for lattice constants in LSDA, PBE, TPSS, and PBEsol.
With the sole and marginal exception of SiC,
LSDA makes lattice constants too short, as indicated by the negative mean
errors.   With the sole and marginal exception of Na,
PBE makes lattice constants too long.  Overall, the systematic PBE overestimate
is close to the systematic LSDA underestimate, as shown by the
total mean absolute errors, and TPSS cures this very little.
On the other hand, PBEsol greatly reduces
this overestimate, by a factor of almost 4, except for
semiconductors, where LSDA is unsurpassed.

\begin{table}[htb]
\begin{tabular}{|c|cccc|}
\hline
error&LSDA&PBE&TPSS&PBEsol\\
\hline
mean error &3.35&0.54&0.18&1.56\\
mean abs. error &3.35&0.67&0.26&1.56\\
\hline
\end{tabular}
\label{t:AE6}
\caption{Errors in atomization energies (eV) for the AE6 
set of molecules, using the 6-311+G(3df,2p) basis set.}
\end{table}
PBEsol is not expected to give good atomization energies.
In Table 2, we give the errors on the AE6 data set of molecules.
These 6 molecules (SiH$_4$, S$_2$, SiO, C$_3$H$_4$ (propyne), 
C$_2$H$_2$O$_2$ (glyoxal), and C$_4$H$_8$ (cyclobutane) were
chosen\cite{LT03}  to be representative, i.e., to reproduce the errors
of much larger data sets.  As is clear, and expected, PBEsol is much
less accurate than PBE, only about halving the error of LSDA.  This can
be largely attributed to PBEsol's worsened total energies of atoms.
For these atoms, the typical error in the total energy per electron
is 0.2-0.3 eV for PBE, but 1.1-1.3 eV for PBEsol.
 
    We have demonstrated the relevance of the second-order
 gradient coefficient for                
the exchange energy of a slowly-varying density to
the bulk and surface properties                 
of solids.  The TPSS meta-GGA \cite{TPSS03}, which incorporates this
 coefficient, gets good surface          
energies but its lattice constants are only marginally
better than those of PBE 
on which it builds, whereas PBEsol is significantly better.
This suggests that an improved meta-GGA might need to recover
the gradient expansion for exchange over a wider range of $\n(\br)$
than TPSS does.

Previous attempts to improve on PBE within the GGA form have retained the PBE
gradient coefficients $\mu$ and $\beta$
for small $s$, but altered the behavior at large
$s$\cite{ZY98,HHN99,WC06,CVSR07}, or have zeroed out
$\mu$\cite{VJKS00,AM05},
and are thus fundamentally different from PBEsol.
The AM05\cite{AM05} functional performs very similarly to PBEsol for the
solids studied here, but 
AM05 follows the proposal of Vitos et al.\cite{VJKS00}
to fit the exchange energy density                                
of an Airy gas (noninteracting electrons in a linear potential, LAG).  
This  is not a                                   
model solid or surface density in any global sense.
This approach does not recover the gradient expansion for
exchange, because the energy density has no such expansion\cite{AM02}.
For $r\s=2$ to $6$, LAG
has errors of 11 to 56\% for $\exsurf$, compared to 1.6 to 4.1\% for
PBEsol.   By retaining the PBE form\cite{PBE96}, PBEsol also correctly
retains the Lieb-Oxford bound and the high-density limit for correlation,
while AM05 does not.  Numerical comparisons and details are available
\cite{EPAPS}.

We have identified the simple exchange-correlation physics underlying
many properties of many solids, and shown how it differs from that for atoms.
We recommend PBEsol for the applications discussed here.
Any existing code that implements PBE can be instantly modified
to try PBEsol, by simply replacing the values of $\mu$ and $\beta$.
Modified PBE subroutines are available from http://dft.uci.edu.
We expect geometries and related
properties to be significantly improved over PBE,
especially for solids under compression.
New pseudopotentials, compatible with PBEsol, must be
generated to use PBEsol with pseudopotential codes.
We thank NSF (CHE-0355405, CHE-0457030, and DMR-0501588) and OTKA for support.

\end{document}